\documentclass[sigconf]{acmart}

\AtBeginDocument{%
  \providecommand\BibTeX{{%
    \normalfont B\kern-0.5em{\scshape i\kern-0.25em b}\kern-0.8em\TeX}}}

\usepackage{mathtools}

\usepackage{amsmath,amsfonts}
\usepackage{graphicx}
\usepackage{textcomp}
\usepackage{multirow}
\usepackage{subcaption}

\usepackage{enumitem}
\usepackage{algorithm}
\usepackage[noend]{algorithmic}
\usepackage{tikz}
\usetikzlibrary{fit,calc}

\newcommand*{\tikzmka}[1]{\tikz[remember picture,overlay,] \node (#1) {};\ignorespaces}
\newcommand{\boxita}[1]{\tikz[remember picture,overlay]{\node[yshift=3pt,fill=#1,opacity=.25,fit={(A)($(B)+(0.118\linewidth,.1\baselineskip)$)}] {};}\ignorespaces}
\newcommand*{\tikzmkb}[1]{\tikz[remember picture,overlay,] \node (#1) {};\ignorespaces}
\newcommand{\boxitb}[1]{\tikz[remember picture,overlay]{\node[yshift=3pt,fill=#1,opacity=.25,fit={(A)($(B)+(0.26\linewidth,.1\baselineskip)$)}] {};}\ignorespaces}
\newcommand*{\tikzmkc}[1]{\tikz[remember picture,overlay,] \node (#1) {};\ignorespaces}
\newcommand{\boxitc}[1]{\tikz[remember picture,overlay]{\node[yshift=3pt,fill=#1,opacity=.25,fit={(A)($(B)+(0.118\linewidth,.1\baselineskip)$)}] {};}\ignorespaces}

\colorlet{blue}{cyan!60}

\usepackage{fancyhdr}

\copyrightyear{2021}
\acmYear{2021}
\setcopyright{usgovmixed}\acmConference[MICRO '21]{MICRO-54: 54th Annual IEEE/ACM International Symposium on Microarchitecture}{October 18--22, 2021}{Virtual Event, Greece}
\acmBooktitle{MICRO-54: 54th Annual IEEE/ACM International Symposium on Microarchitecture (MICRO '21), October 18--22, 2021, Virtual Event, Greece}
\acmPrice{15.00}
\acmDOI{10.1145/3466752.3480113}
\acmISBN{978-1-4503-8557-2/21/10}

\begin{document}
\title{I-GCN: A Graph Convolutional Network Accelerator with Runtime Locality Enhancement through Islandization } 
\settopmatter{authorsperrow=1}
\author{Tong Geng$^\dag$, Chunshu Wu$^\ddag$, Yongan Zhang$^\S$, Cheng Tan$^\dag$, Chenhao Xie$^\dag$, Haoran You$^\S$, Martin C. Herbordt$^\ddag$, Yingyan Lin$^\S$, Ang Li$^\dag$}

\affiliation{%
\vspace*{0.15truein}
  \institution{$\dag$ Pacific Northwest National Laboratory, Richland, WA}
  \institution{$\ddag$ Boston University, Boston, MA}
  \institution{$\S$ Rice University, Houston, TX\\
 $\{$tong.geng, cheng.tan, chenhao.xie, ang.li$\}$@pnnl.gov,$\{$happycwu, herbordt$\}$@bu.edu,$\{$yz87,hy34,yingyan.lin$\}$@rice.edu}
 \vspace*{0.2truein}
 \country{}
  }

\renewcommand{\shortauthors}{Tong Geng, et al.}



\begin{abstract}
    Graph Convolutional Networks (GCNs) have drawn tremendous attention in the past three years. 
    Compared with other deep learning modalities, high-performance hardware acceleration of GCNs is as critical but even more challenging. The hurdles arise from the poor data locality and redundant computation due to the large size, high sparsity, and irregular non-zero distribution of real-world graphs.
    
    In this paper we propose a novel hardware accelerator for GCN inference, called I-GCN, that significantly improves data locality and reduces unnecessary computation. The mechanism is a new online graph restructuring algorithm we refer to as {\it islandization}. The proposed algorithm finds clusters of nodes with strong internal but weak external connections. The islandization process yields two major benefits. First, by processing islands rather than individual nodes, there is better on-chip data reuse and fewer off-chip memory accesses. Second, there is less redundant computation as aggregation for common/shared neighbors in an island can be reused. The parallel search, identification, and leverage of graph islands are all handled purely in hardware at runtime working in an incremental pipeline. This is done without any preprocessing of the graph data or adjustment of the GCN model structure.  
    Experimental results show that I-GCN can significantly reduce off-chip accesses and prune \textcolor{black}{38\%} of aggregation operations, leading to performance speedups over CPUs, GPUs, the prior art GCN accelerators of \textcolor{black}{5549$\times$, 403$\times$, and 5.7$\times$} on average, respectively.

\end{abstract}



\maketitle

\section{Introduction}

Conventional deep learning paradigms such as Convolution Neural Networks (CNNs) \cite{krizhevsky2012imagenet} and Recurrent Neural Networks (RNNs) \cite{mikolov2010recurrent} have been demonstrated to be quite efficient, but primarily for applications using Euclidean data \cite{wu2020comprehensive,Geng21,Wang20,o3bnn}. Many other applications, however, require that the relationships between data points be conserved and must therefore use non-Euclidean data structures such as graphs \cite{liu2020guiding, coley2019graph, xie2018crystal, zitnik2018modeling, yang2019aligraph, nguyen2018iot}. To fill this need, Graph Neural Networks (GNN) have been proposed \cite{bruna2013spectral,defferrard2016convolutional,kipf2016semi, yun2019graph}.

Graph Convolutional Networks (GCNs) are a type of GNN that has drawn tremendous attention in the past three years due to their unique ability to extract latent information from graph data. Practical applications of GCNs include prediction of cascading power-grid failure \cite{liu2020guiding}, E-commerce \cite{yang2019aligraph}, and etc \cite{nguyen2018iot,coley2019graph}. The deployment of GCNs in these applications typically poses strict constraints on latency and throughput.

To satisfy the increasingly stringent performance requirements, designing high-performance hardware accelerators for GCNs becomes urgent \cite{yan2020hygcn, geng2019uwb}. Real world graphs tend to have large size, high sparsity, and extremely unbalanced non-zero distributions; therefore, the direct application of existing methods, such as Sparse CNNs \cite{han2016eie, zhang2016cambricon, kim2017novel}, has been reported to be insufficient \cite{xie2014distributed, gonzalez2012powergraph, abou2006multilevel, latapy2008main}.

We briefly discuss the performance challenges of the two major graph aggregation methods used in GCN acceleration:

\vspace{4pt}\noindent
(1) In \textbf{PULL-based aggregation} nodes are evaluated sequentially, but for each node: the neighbor features are gathered (i.e., pulled) simultaneously for aggregation. The advantage of the {\it pull} method is that since nodes are processed in order, a small buffer is sufficient to accommodate the aggregation results. In other words, it shows good reuse for the result matrix. The major problem, however, is the \emph{poor data locality for accessing the feature matrix}. Given that the adjacency matrix of real-world graphs is typically very sparse and imbalanced, parallel fetches of the corresponding neighbor features can be random and non-coalesced. Since the feature matrix can be too large to fit into on-chip memory, repeated irregular off-chip data accesses for the feature matrix are required; this process is bounded by off-chip bandwidth. HyGCN \cite{yan2020hygcn} uses PULL approach. Although data-aware sparsity-elimination hardware is used to reduce off-chip data accesses, feature matrices still need to be accessed many times. An HBM is required to avoid hardware starvation.

\vspace{4pt}\noindent
(2) In \textbf{PUSH-based aggregation} nodes are evaluated in parallel, but the feature data of the nodes are distributed sequentially (i.e. pushed) to their neighbors for aggregation. This avoids the irregularity in accessing the feature matrix, but essentially shifts the burden to the result matrix. Given that aggregation for a node is processed sequentially, a large buffer is required to hold the partial results. If there are too many nodes, the partial result buffer cannot fit into on-chip memory, leading to frequent off-chip access and bandwidth saturation. Furthermore, the power-law distribution of non-zeros in the adjacency matrix can cause serious workload imbalance. AWB-GCN \cite{geng2019uwb} practices the PUSH approach and uses a run-time workload autotuning architecture to effectively tackle the workload imbalance problem. However, it does not address the data locality problem for the irregular and random accesses of the result matrix, which can be the most critical problem in GNN processing especially for large graphs. 

\vspace{4pt}
Clearly, the key problem is the irregularity (i.e., distribution of non-zeros) of the adjacency matrix, which leads to poor data reuse in accessing either the feature matrix or the result matrix. A recent trend therefore is to rely on offline preprocessing to restructure the adjacency matrix to improve data locality, e.g. in Rubik \cite{chen2020rubik} and GraphACT \cite{zeng2020graphact}. The implicit assumption is that the graph structure (i.e., the adjacency matrix) is fixed upon inference so that the large overhead of graph restructuring can be omitted in the critical path. However, this is not always the case, as real-world graphs are frequently updated (e.g., evolving graphs) or generated dynamically (e.g., inductive graphs). The high restructuring overhead, e.g., seconds in Rubik and GraphACT, is not tolerable when processed online. Besides, although both Rubik and GraphACT allow for the possibility of computation reuse for shared neighbors during aggregation, their complex software-based reordering algorithms introduce significant delay and are only feasible when processed offline.

\begin{figure}[t] 
\centering
\includegraphics[width=3.2in]{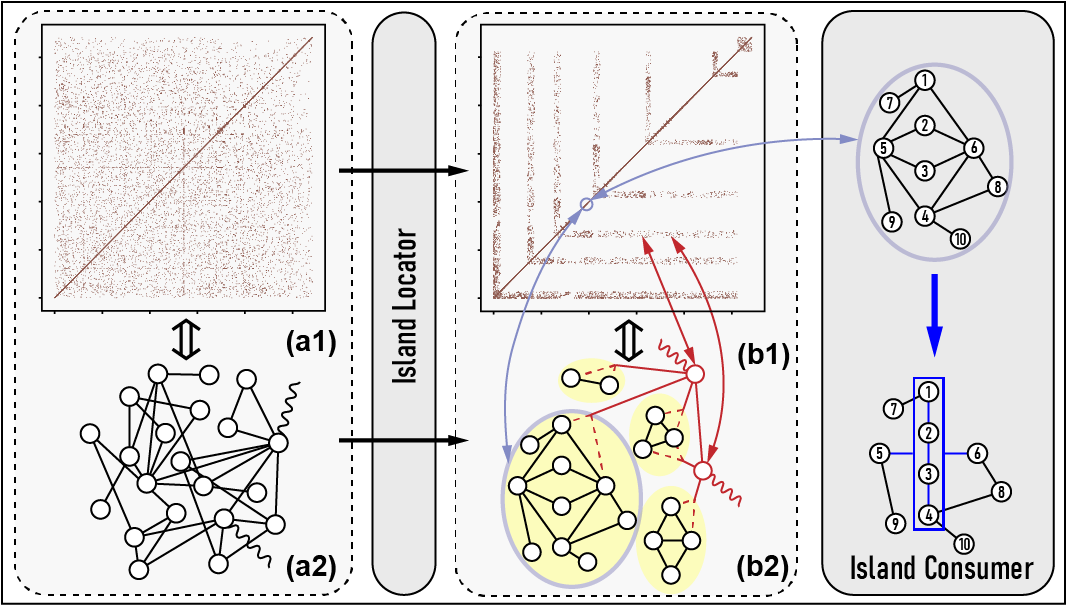}
\vspace{-0.10truein}
\caption{Workflow of I-GCN.}
\label{big_picutre}
\vspace{-0.15truein}
\end{figure}

In this paper, we propose a novel hardware accelerator, called I-GCN, which implements a new online graph restructuring algorithm -- {\it islandization} -- that can significantly improve data locality and reduce redundant computation for GCN inference acceleration.
Specifically, I-GCN's {\it Island Locator} module, at runtime, is able to detect the {\it hub} nodes (i.e. nodes with high degree), find their neighbors, and then use these neighbor nodes as starting points to explore and locate islands iteratively. {\it Islands} are groups of nodes with strong internal, but no external, connections other than with hubs. Note that islands often (but not always) have practical semantics: in a social network they might correspond to people working in the same institute; in a citation network they might correspond to papers published in the same conference series. 
Determining islands is non-obvious, especially in typical (huge and sparse) adjacency matrices.

Figure~\ref{big_picutre} gives an overview of I-GCN. \textbf{(i) Island Locator:} after identifying the islands, the non-zeros of the adjacency matrix become highly clustered -- the none-zeros of hubs and islands form the L-shapes and the anti-diagonal respectively. \textbf{(ii) Island Consumer:} using the hub and island information, the Island Consumer performs aggregation and combination in a fine-grained pipelined manner. Redundant aggregation is skipped. This process continues until all nodes are determined to be either hubs or islands. The benefits of islandization are two-fold: (1) \emph{Improving on-chip data locality.} Through clustering, accesses to the feature and result matrices can be constrained within a much smaller working-set (i.e., each L-shape and island in Figure~\ref{big_picutre}). This greatly improves on-chip data reuse and avoids a tremendous number of off-chip accesses. (2) \emph{Reducing redundant computation.} After clustering, nodes within a cluster tend to share a large portion of common neighbors. During GCN aggregation, rather than repeatedly counting each neighbor, aggregated information about the common neighbors as a whole can be distributed and reused, avoiding repeated aggregation calculation for common neighbors. This neighbor-sharing information is mostly ambiguous in the raw adjacency matrix, but becomes obvious after islandization (see node 1$\sim$4 in Figure~\ref{big_picutre}). 
In summary, islandization resolves the locality issues in both Pull and Push approaches. 

This is the first work, to the best of our knowledge, that tackles the fundamental data locality problem of GCN acceleration and efficiently skips redundant aggregation through online hardware-based graph restructuring. This paper makes the following contributions:

\begin{itemize}[leftmargin=.11in]
\item We propose a novel islandization algorithm for efficient runtime parallel graph restructuring, which can significantly improve on-chip data locality.   
\item We design a new hardware accelerator architecture called I-GCN that effectively implements the islandization algorithm, harvesting the data locality exposed through islandization and avoiding redundant aggregation among shared neighbors.
\item Experimental results show that I-GCN can significantly reduce off-chip accesses and prune \textcolor{black}{38\%} of aggregation operations, leading to performance speedups over PyG- \& DGL-based CPUs, PyG- \& DGL-based GPUs, and prior art GCN accelerators of \textcolor{black}{9568$\times$ \& 1243$\times$, 368$\times$ \& 453$\times$, and 5.7$\times$}, respectively.
\end{itemize}

\section{Background and Motivation}

\begin{figure*}[!th] 
\centering
\includegraphics[width=7in]{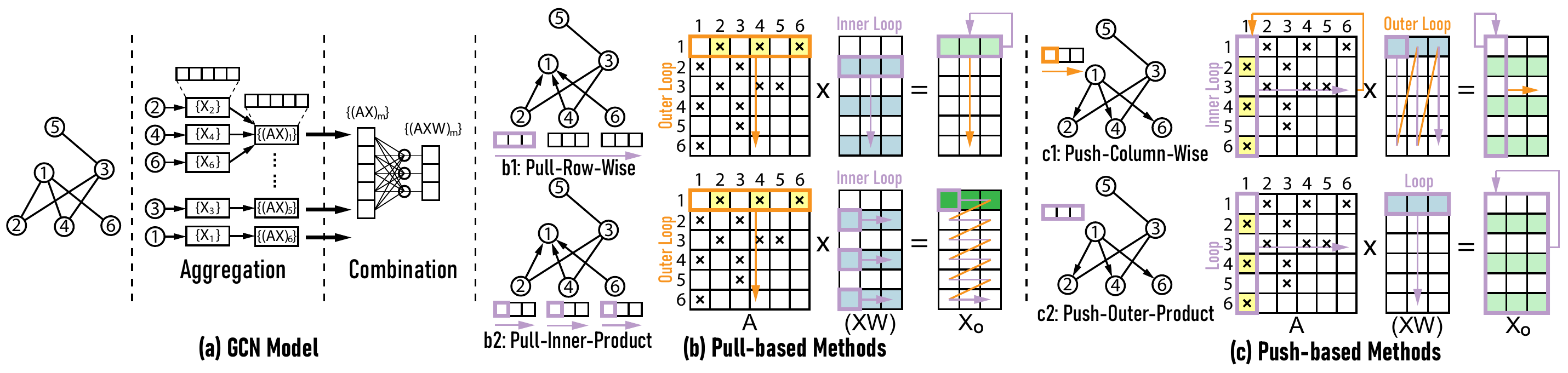}
\vspace{-0.25truein}
\caption{\textcolor{black}{PULL-based (Row-wise \& Inner-Product) and PUSH-based (Column-wise \& Outer-Product) methods.}}
\vspace{-0.0truein}
\label{gcn_pull_push}
\end{figure*}

We first introduce GCN algorithms and then elaborate the existing GCN processing methodologies and their challenges.

\subsection{Graph Convolutional Networks}

GCNs are composed of stacked GraphCONV layers. The computation flow of each GraphCONV layer includes two phases: \emph{Aggregation} and \emph{Combination}, as illustrated in Figure~\ref{gcn_pull_push}-(a). In the aggregation phase, each node gathers and aggregates features of its neighbor nodes to update the local feature-vector. In the combination phase, the updated feature-vectors are further merged to extract high-level abstraction through a local Multi-Layer Perceptron (MLP) network. From the perspective of linear algebra, the layer-wise forward propagation of GCN can be expressed as Equation~\ref{eq:gcn_layer}:

\begin{equation}
X^{(l+1)} = \sigma(\tilde{A} X^{(l)}  W^{(l)})
\label{eq:gcn_layer}
\end{equation}

where $\tilde{A}$ is the graph adjacency matrix. $X^{(l)}$ is the matrix of input features in layer-$l$; $W^{l}$ is the weight matrix of layer-$l$. $\sigma(.)$ denotes non-linear activation functions \cite{krizhevsky2012imagenet}. 

There are many modalities beneath the umbrella of GCNs, e.g. GraphSage \cite{hamilton2017inductive,chen2018fastgcn} and Graph Isomorphism Network (GIN) \cite{xu2018how}. As shown in \cite{GCNAX}, the forward propagation of most GCNs can be abstracted and expressed as Equation~\ref{eq:gcn_layer}.

\subsection{\textcolor{black}{Design Space Exploration}}

We first discuss the design choices related to execution order of the two phases in GraphCONV. We then present the design choices within each phase.

\subsubsection{Execution Order of a GraphCONV Layer}

There are two alternative computation orders for a GraphCONV layer: \emph{aggregation first} ($(AX)\times W$) and \emph{combination first} ($A\times (XW)$). Existing work \cite{geng2019uwb,liang2020engn} has shown that the combination first approach can reuse the same Sparse-dense Matrix Multiplication (SpMM) kernel for both multiplications $XW$ and $A(XW)$ and incorporates less arithmetic computation. I-GCN follows this combination first approach.

\subsubsection{Aggregation Phase}

We first discuss the design choices of the aggregation phase, which is normally the performance bottleneck of GCN processing. There are two typical methods of aggregation: PULL and PUSH. For clarity, we compare them from a linear algebra perspective as illustrated in Figure~\ref{gcn_pull_push}.
We discuss the correspondence between PULL/PUSH-based graph computation methods and four SpMM approaches as summarized in \cite{srivastava2020matraptor}: inner-product, outer-product, row-wise, and col-wise.

PULL-based methods aggregate nodes sequentially. For each particular node, the feature vectors of all its neighbors are gathered and aggregated in two ways: (1) PULL-Inner-Product: features from different channels are calculated sequentially, analogous to the inner-product approach of SpMM. (2) PULL-Row-Wise: features from all channels are calculated in parallel, similar to the row-wise-product approach of SpMM.
As shown in Figure~\ref{gcn_pull_push}-(b), PULL-based methods always process non-zeros of matrix A and produce the aggregation result matrix $X_o$ by rows (\textit{\textbf{outer loop}}). To calculate each row of $X_o$ (\textit{\textbf{inner loop}}), PULL-Row-Wise (Figure~\ref{gcn_pull_push}-(b1)) fetches the entire feature vectors (i.e. entire rows of $XW$) of required nodes and performs vector accumulation sequentially; in contrast, PULL-Inner-Product fetches features by channel (i.e. column of $XW$) and computes the output features in $X_o$ sequentially (Figure~\ref{gcn_pull_push}-(b2)).

PULL-based methods have their advantages and disadvantages. On the plus side, they reuse matrix A and only require relatively small on-chip buffers to conserve the partial aggregation results (since they are produced row by row). However, they both suffer from poor data reuse of matrix $XW$. Due to the scattered and irregular distribution of non-zeros in matrix A, the rows to be accessed in XW can be random. Given that the height of XW equals the number of nodes in the graph, which can be very large, XW (which is dense) can rarely be stored on-chip, leading to frequent data fetches and inferior performance. Ideally, if the non-zeros from the same column of A can be clustered, they can reuse the same row of XW, improving data locality. 

PULL-Row-Wise is more popular than PULL-Inner-Product in prior art mainly due to: (1) accessing the entire rows is more efficient than randomly fetching elements from different columns in terms of off-chip access; (2) as $XW$ is dense, the processing of a row in a fixed size can be parallelized without introducing workload imbalance.

\textcolor{black}{PUSH-based aggregation, in contrast, calculates the aggregated features of all nodes simultaneously by broadcasting features of each node to all its neighbors one after another. Once a node receives the features from its neighbors, it updates its local feature vector. There are two ways of feature broadcasting: PUSH-Column-Wise and PUSH-Outer-Product. With PUSH-Column-Wise, the input features are broadcast by channel; the output features are calculated by channel. As shown in Figure~\ref{gcn_pull_push}-(c1), at iteration $k$ of \textit{\textbf{outer loop}}, each node only pushes its features at channel k (i.e. column k of XW) to the neighbors. Once all nodes have broadcast their features at channel k to the neighbors (\textit{\textbf{inner loop}}), every node has the complete aggregated features at channel k (i.e. column k of $X_o$). In contrast, PUSH-Outer-Product (Figure~\ref{gcn_pull_push}-(c2)) method processes all channels at the same time. The entire input feature vector of each node is broadcast to all its neighbors in parallel. Once done, the feature aggregation of the entire graph finishes. In the matrix perspective, PUSH-Outer-Product processes the non-zeros of matrix A by column, accesses the features of XW by row, and updates the corresponding partial results of $X_o$ with respect to the row IDs of non-zeros in matrix A.}

\textcolor{black}{The advantage of PUSH-based methods over PULL is that the data of XW can be fully reused via feature broadcasting. However, the accesses to $X_o$ then become scattered and random. Given the height of $X_o$ is equal to the number of nodes in the graph, which can be large, even a single column of $X_o$ may not be able to be buffered on-chip. Frequent accesses to $X_o$ thus render repeated data fetches from off-chip DRAM. Ideally, if the non-zeros from the same row of A can be clustered, they can reuse the same row of $X_o$, improving data locality and performance. }

\textcolor{black}{PUSH-Column-Wise method is more popular than PUSH-Outer-Product in prior art because for graphs that a single column of $X_o$ can be put on-chip, 
PUSH-Column-Wise method can avoid the repeated off-chip data accesses over $X_o$. However, this does not work for large graphs. Consequently, PUSH-Column-Wise does not fundamentally handle the data locality issue. Additionally, it needs to repeatedly access matrix A which also incurs additional off-chip access.}

\begin{table}[t]
\setlength{\tabcolsep}{1.5pt}
\footnotesize
\centering
\begin{tabular}{|c|c|c|c|c|c|c|c|}
\hline
                & \begin{tabular}[c]{@{}c@{}}On-chip\\ Storage\end{tabular} & \begin{tabular}[c]{@{}c@{}}Off-chip\\ Access\end{tabular} & \begin{tabular}[c]{@{}c@{}}Reuse\\ $XW$\end{tabular} & \begin{tabular}[c]{@{}c@{}}Reuse\\ $A$\end{tabular} & \begin{tabular}[c]{@{}c@{}}Reuse\\ $X_o$\end{tabular} & \begin{tabular}[c]{@{}c@{}}Load\\ Imbalance\end{tabular} & \begin{tabular}[c]{@{}c@{}}Redundancy\\ Removal\end{tabular} \\ \hline
PULL            & Low                                                         & High                                                            & Low                                                       & High                                                      & High                                                           & No                                                    & Hard        \\ \hline
PUSH            & High                                                         & High                                                            & High                                                       & Low                                                      & Low                                                           & Yes                                                 & Hard             \\ \hline
Islandization & Low                                                         & Low                                                            & High                                                       & High                                                      & High                                                           & No                                                 & Easy             \\ \hline
\end{tabular}
\vspace{0.10truein}
\caption{Comparison of PULL, PUSH, Island methods.}
\vspace{-0.3truein}
\label{tb:pull_push}
\end{table}

Table~\ref{tb:pull_push} summarizes the advantages and disadvantages of both approaches. The proposed islandization method is capable of overcoming all the drawbacks by clustering the non-zeros at runtime and provides nearly optimal data reuse in GCN inference. Furthermore, with islandization, the unnecessary aggregation for commonly shared neighbors can be identified much more easily for effective pruning.

\subsubsection{Combination Phase}

As both combination and aggregation are based on SpMM kernels, the combination phase shares similar issues with aggregation phase. The only difference is that the weight matrix W in combination is normally much smaller than the feature matrix XW in aggregation, and can be more likely stored on-chip. Therefore, the data locality issue of the PULL-based method is less prominent than in aggregation. Consequently, I-GCN adopts the PULL-based method for combination.

\section{Algorithms and Architectures}

This section first presents the overall workflow and hardware architecture of I-GCN. It then introduces the algorithms and the corresponding architectures of two major components of I-GCN: the \emph{Island Locator} and the \emph{Island Consumer}.

\subsection{I-GCN Overview}

\subsubsection{I-GCN Workflow}

\begin{figure}[!t] 
\centering
\includegraphics[width=3.4in]{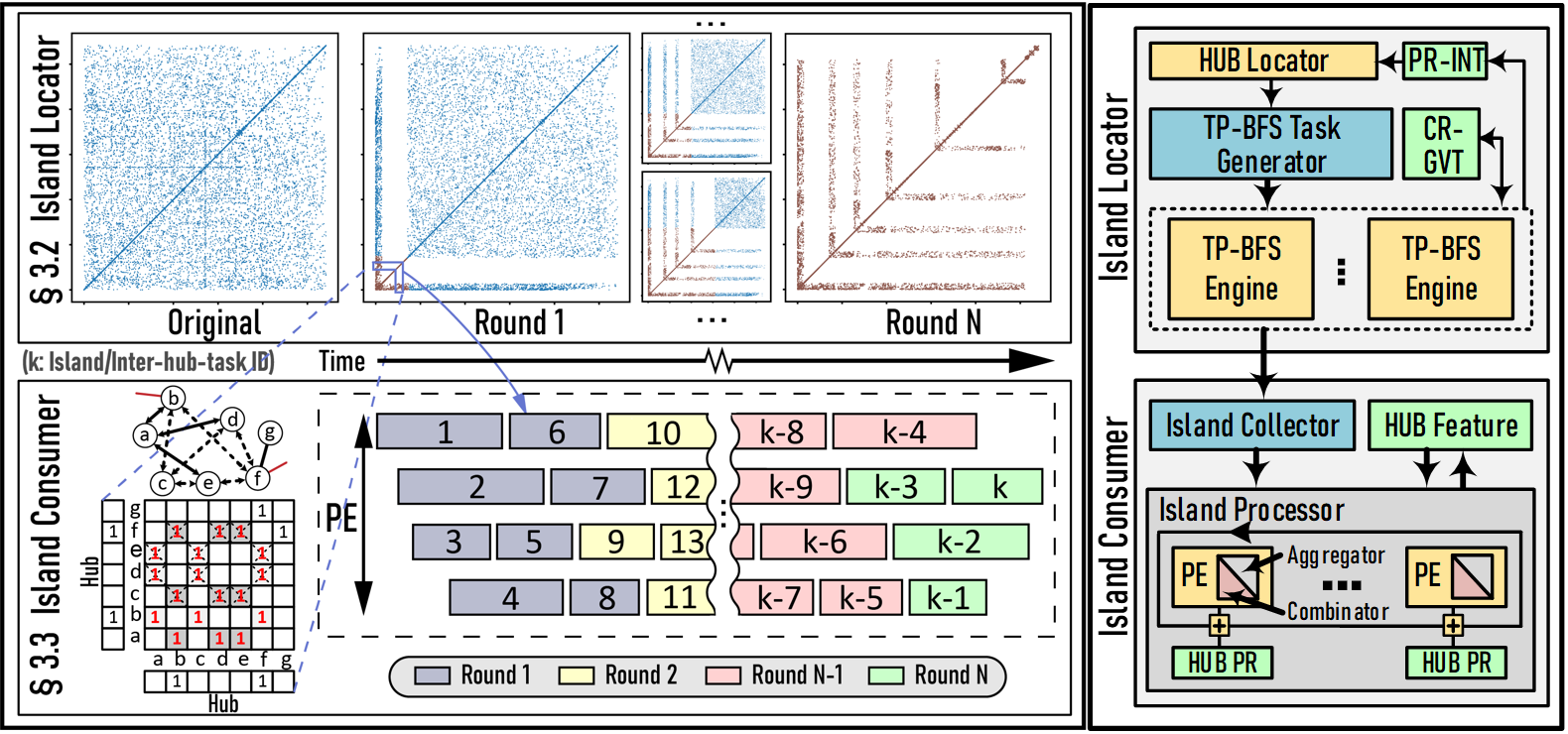}
\vspace{-0.15truein}
\caption{\textcolor{black}{Detailed I-GCN workflow and overview of I-GCN architecture}}
\vspace{-0.15truein}
\label{design_overview}
\end{figure}

Figure~\ref{design_overview} illustrates how I-GCN improves data locality through islandization: the clustering of graph nodes into islands. Elements of an island tend to have strong internal connections, and are connected to other islands through \textbf{\textit{hubs}} -- \textcolor{black}{nodes with high fan-in/out degrees, which act as the points of contact for islands.} Islandization is the process of finding these structures hidden in the original graph. At high level, I-GCN processing begins by locating these islands, after which it processes the graph at the granularity of hubs and islands rather than nodes, thus being able to improve locality and reduce redundant computation.

I-GCN creates the islands gradually by round. Figure~\ref{design_overview} shows the process using the CORA dataset. The scattered non-zeros belonging to the same island are clustered around the diagonal. The islands (including the hubs and nodes) are discovered in parallel by a hardware module called \textbf{Island Locator}. In each round, the Island Locator uses different thresholds to recognize hubs and then shapes the island by scanning the hubs' neighbor nodes. As shown in Figure~\ref{design_overview}, the processing moves from the bottom-left corner of the adjacency matrix to the upper-right corner, along the anti-diagonal. In each round a portion of the non-zeros are gathered into an L-shape cluster (except the non-zeros around the anti-diagonal), leaving the remaining non-zeros untouched. With new hubs located at each round, new L-shape clusters are formed. This process continues until all non-zeros are clustered either into L-shapes (hubs) or around the anti-diagonal (islands) (see \emph{Round N} in Figure~\ref{design_overview}). 

In parallel, whenever an island is formed its adjacency information is forwarded to the second module called \textbf{Island Consumer}. Island Consumer then processes the island as a small but dense sub-graph, fetches its node features, and performs the required combination and aggregation. 
Note that although the Island Locator forms the islands gradually by round, and requires per round synchronization, the Processing Elements in the Island Consumer can process an island as soon as it is formed without synchronization at the completion of each round: it does not need to wait until all the islands in an islandization round are formulated. I-GCN overlaps graph restructuring and graph processing.

As shown in Figure~\ref{design_overview}, each node in an island, labeled as an island node, only connects to the nodes of the same island and the hub nodes connected to the island. This ensures that the space between the L-shapes is purely blank.
Therefore, when processing a GraphCONV layer, the associated data of island nodes are only needed when the island is being processed. Consequently, they only need to be fetched from off-chip once. The hubs do have the chance of being used multiple times during the processing of different islands and inter-hub connections. However, since hubs are normally a small fraction of the entire graph, their associated data will likely be stored on-chip and sufficiently reused. Even if the hubs' associated data is too large to fit in the on-chip memory, our method still reduces off-chip data movement. 

In summary, for GraphCONV processing of real-world graphs with component structures using I-GCN, most data are fetched only once, except the adjacency data of some island nodes which may need to be accessed multiple times during the multi-round island locating. We evaluate this in Section~4.6. Note that component structures are commonly observed in real-world graphs.


Another benefit of islandization is that it assembles the redundant aggregation among shared neighbors in the processing of each island. These redundant operations, originally hidden in the large graph, are illuminated after islandization. 

As shown in Figure~\ref{design_overview}, islands are formed by nodes with intensive internal links, most of which are among shared neighbors. During the processing of the small and dense islands, it becomes easy to recognize these repeated computations and skip them. As an example, Figure~\ref{design_overview}(A) shows the sub-graph structure of the sixth island found in the first round. The bitmap at the lower-left corner is the purple block in the Round 1 adjacency matrix enlarged. The two hub vectors are from the L-shape, while the 7 $\times$ 7 matrix includes the adjacency data of the island nodes. During the evaluation of the island, Island Consumer will find the redundant aggregation operations for the shared neighbors and skip them. The redundancy removal methodology is discussed in detail in Section 3.3. 

\subsubsection{I-GCN Overall Architecture}

The overall architecture of I-GCN, including  Island Locator and Island Consumer, is shown in Figure~\ref{design_overview}(B). The HUB Locator in the Island Detector is responsible for recognizing hubs. The hubs found are forwarded to TP-BFS Task Generator. TP-BFS is short for \textbf{Threshold-based and Parallel Breadth-First Search}. Task Generator will generate and assign BFS tasks for islandization. These tasks are conducted by TP-BFS Engines. Once TP-BFS locates an island, the related adjacency and node ID information are forwarded to Island Collector in Island Consumer. Island Collector distributes the island information to an idle PE for performing its combination and aggregation jobs. Furthermore, Island Collector also generates and distributes new tasks which include inter-hub connections based on the hub information collected by TP-BFS engines at each round. Detailed architectures are presented next.

\subsection{Island Locator}

We first introduce the algorithm used in the Island Locator and then present its architectural support. For clarity we use a homemade graph (Figure~\ref{fig:toy_example}) to illustrate the process.

\subsubsection{Algorithm}

\begin{figure}[t!]
\vspace{-0.07truein}%
    \begin{minipage}{0.47\textwidth}
        \begin{algorithm}[H]
        \small
          \caption{\textbf{Island Locator Algorithm:} Th1, Th2, and Th3 are executed concurrently and asynchronously.}
           \label{al:1}
        \begin{algorithmic}[1]
         \STATE {\bfseries Inputs:} $N$: Node list of the input graph; $(P1, P2)$: parallel factors for hub detection and island searching; $TH_{o}$: initial threshold for hub nodes; $c_{max}$: the maximum number of nodes that an island can have; Decay(): hub threshold decay function
         \STATE {\bfseries Outputs:} $l_{islands}$: list of islands' nodes and hubs info 
          \STATE $TH_{tmp} = TH_{o}$; $l_{islands} = \{\}$
          \WHILE{$|N| > 0$}
            \STATE $task =\{\}$;  $hub\_buffer =\{\} $
            \\\textcolor{purple}{\scriptsize \# pop hub nodes from graph $N$ to $hub\_buffer$}
            \STATE \tikzmka{A} Th1: \textbf{detect$\_$hub}($N$, $P1$, $TH_{temp}$, $hub\_buffer$)\tikzmka{B} \boxita{blue} \vspace{0.0em}
            \\\textcolor{purple}{\scriptsize \# pop neighbors of nodes in $hub\_buffer$ to $task$}
            \STATE \tikzmkb{A} Th2: \textbf{task\_assign}($hub\_buffer$, $task$)\tikzmkb{B} \boxitb{red} \vspace{0.0em}
            \\\textcolor{purple}{\scriptsize \# explore islands from nodes in $task$ using BFS}
            \STATE \tikzmkc{A} Th3: \textbf{TP-BFS}($task$, $P2$, $TH_{temp}$, $c_{max}$,  $l_{islands}$) \tikzmkc{B} \boxitc{orange} 
            \STATE{\bfseries [ Hold until parallel Th1/2/3 finish]}
            \STATE $TH_{tmp} =$ Decay($TH_{tmp}$)  
          \ENDWHILE
          \RETURN $l_{islands}$
        \end{algorithmic}
        \end{algorithm}
    \end{minipage}
\end{figure}

\begin{figure}[t!]
\vspace{-0.20truein}%
    \begin{minipage}{0.47\textwidth}
        \begin{algorithm}[H]
        \small
          \caption{ \textbf{detect$\_$hub}: sweep nodes and move nodes with degrees larger than thresholds to container $hub\_buffer$}
           \label{al:2}
        \begin{algorithmic}[1]
         \STATE {\bfseries Inputs:} $N$: input graph node list; $P1$: parallel factor; $TH_{tmp}$: threshold for hub's degree; $hub\_buffer$: container of the hub nodes
         \STATE {\bfseries Outputs:} None (modify $hub\_buffer$ in place )
            \FOR{ $b=0$ {\bfseries to} $\lceil|N|/P1\rceil$ } 
                \FOR{$p=0$ {\bfseries to}  $P1$ {\bfseries in parallel}}
                    \STATE  Check node $N[b*P1+p]$ as $n_o$
                    \IF{$n_o \in l_{islands}$}
                        \STATE Pop $n_o$ from $N$
                    \ELSIF{$n_o.degree \geq TH_{tmp} $}
                            \STATE Pop $n_o$ from $N$ to $hub\_buffer$
                    \ENDIF
                \ENDFOR 
            \ENDFOR
        \end{algorithmic}
        \end{algorithm}
    \end{minipage}
\end{figure}

\begin{figure}[t!]
\vspace{-0.20truein}%
    \begin{minipage}{0.47\textwidth}
        \begin{algorithm}[H]
            \small
          \caption{ \textbf{task\_assign}: pop nodes from $hub\_buffer$ and add them and their neighbors to task queue}
           \label{al:3}
        \begin{algorithmic}[1]
         \STATE {\bfseries Inputs:} $hub\_buffer$: container of the hub nodes; $task$: container of nodes to be chosen as potential starting points for BFS 
         \STATE {\bfseries Outputs:} None (modify $task$ in place)
            \WHILE{$|hub\_buffer| > 0$}
                \STATE Pop a node from $hub\_buffer$  
                \STATE Append the popped hub node and each of its neighbors to $task$, in the form of tuples.
            \ENDWHILE
        \end{algorithmic}
        \end{algorithm}
    \end{minipage}
    \vspace{-0.10truein}
\end{figure}

\begin{figure}[t] 
\centering
\vspace{-0.05truein}%
\includegraphics[width=3.3in]{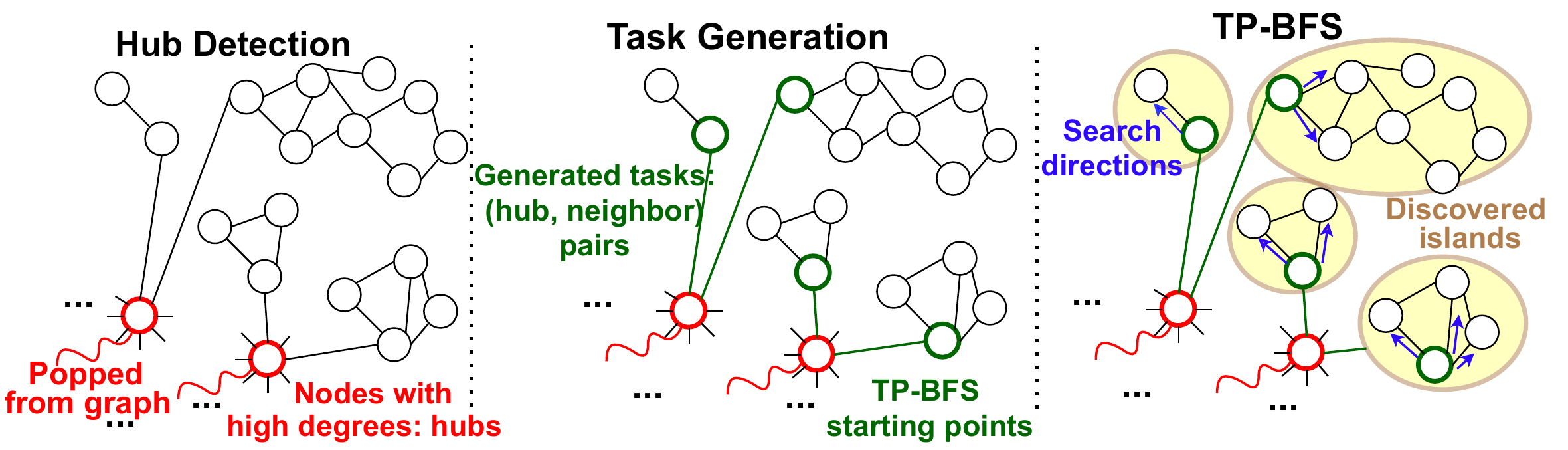}
\vspace{-0.07truein}
\caption{Toy example of Island Locator algorithm.}
\label{fig:toy_example}
\vspace{-0.15truein}
\end{figure}

Algorithm~\ref{al:1} describes the simplified workflow of the Island Locator; \textcolor{black}{a toy example can be found in Figure~\ref{fig:toy_example}.} \textcolor{black}{Overall the purpose of the algorithm is to locate the hubs and use their neighbors as starting points to search for islands (with the \textbf{TP-BFS} algorithm). To boost parallelism}, the algorithm comprises three concurrent asynchronous tasks: \emph{hub detection} \textcolor{black}{(line 6)}, \emph{BFS task generation} \textcolor{black}{(line 7)}, and \emph{TP-BFS execution} \textcolor{black}{(line 8)}. Hub detection and TP-BFS are additionally performed in parallel \textcolor{black}{across the $P1$ and $P2$ \textbf{parallel for loops} in algorithm 2 (line4) and 4 (line5).}
As mentioned in Section 3.1, locating islands is performed by rounds, \textcolor{black}{which are iterations of line 4.} \textcolor{black}{$TH_{tmp}$ represents the most current hub detection threshold, which is modified at run-time.} Note that synchronization is required among the three tasks at the start of each round \textcolor{black}{(line 9)}. Specifically, the algorithm takes the input of (1) node list $N$; (2) parallel factors \textcolor{black}{$P1$ and $P2$}, which define the numbers of the parallel FIFOs and TP-BFS engines, respectively;
(3) the initial hub threshold $TH_o$, which marks the initial minimum degree of the hub nodes; (4) the maximum number of nodes in an island, $c_{max}$; and (5) Decay(), which defines the function to decrease the hub detection threshold.
The algorithm then outputs the abstract container $l_{islands}$, which encapsulates all island-related information. This includes island nodes' indices and neighbors, the connected hub nodes' indices and their connections with island nodes, the number of island and hub nodes, and etc. Note that this abstract container is used just for clarity.

Island location starts with parallel hub detection. This \textcolor{black}{is shown in Algorithm~\ref{al:2}, which uses a threshold-based method to find hubs by rounds. If the degree of a node is above the current threshold $TH_{tmp}$, the node is identified as a hub and inserted into a container(buffer) $hub\_buffer$. $TH_{tmp}$ is reduced each round (line 10 of Algorithm \ref{al:1})} until all nodes are classified as island or hub nodes. To accelerate parallel hub detection, in each round we remove the nodes already classified as hub and island nodes in the previous round. At the end of island location, the node list $N$ should be empty. 

Hub detection is followed by BFS task generation (Algorithm \ref{al:3}). Once a hub node is detected, the Island Locator finds its neighbors by accessing its adjacency list and caches these neighbor nodes in a task queue, $task$. \textcolor{black}{The Island Locator sends neighbor nodes to the task queue, $task$, and then to parallel BFS engines where the nodes are used by TP-BFS as the starting points for forming islands. Here we use neighbor nodes as the starting nodes (instead of hubs) in order to extract higher parallelism from TP-BFS. This significantly improves the scalability of the Island Locator.}

Once the first task is generated and stored in the task queue, TP-BFS starts (see Algorithm~\ref{al:4}). \textcolor{black}{The algorithm keeps track of the number of nodes whose neighbors have been explored exhaustively ($query$) and the total number of visited nodes ($count$). Once $query$ catches up with $count$, an island is found with all the nodes and their neighbors within the island completely explored and the island information is returned.} Multiple TP-BFS engines are able to work in parallel on different neighbors of different hubs. \textcolor{black}{Multiple $v_{local}$s are used to keep track of the nodes visited locally by each TP-BFS engine while $v_{global}$ is used to keep track of the nodes visited by any of the TP-BFS engines.}

\textcolor{black}{Each engine begins island searching by (a) recording the initial node, $a_o$, as the first visited node in a locally visited list, $v_{local}$; (b) setting the $query$ pointer to zero indicating no node's neighbors have been fully explored; and (c) accessing the node's neighbors. Note that if an engine finds $a_o$ also a hub, it will drop the task and forward this inter-hub connection information to Island Collector.}

As the Island Locator overlaps hub detection and TP-BFS, the TP-BFS engine does not know which nodes are the hubs and should not be considered as island nodes. Therefore, when the neighbors of the first node arrive, the engine checks whether it is a hub node \textcolor{black}{(line 11 of Algorithm \ref{al:4}).} In addition, the engine must also check whether it has itself visited this node during the execution of the current task \textcolor{black}{(line 12 of Algorithm \ref{al:4}).} When either happens, the engine needs to skip the node and work on the next neighbor. Otherwise, the engine would have appended the node to the local visited list and increment the $counter$ \textcolor{black}{(line 16 of Algorithm \ref{al:4}).}

\begin{figure}[t!]
\vspace{-0.07truein}%
    \begin{minipage}{0.47\textwidth}
        \begin{algorithm}[H]
        \small
          \caption{ \textbf{Simplified TP-BFS}: fetch nodes from the task queue and use them as starting points in island detection}
           \label{al:4}
        \begin{algorithmic}[1]
         \STATE {\bfseries Inputs:} $task$: container of nodes to be chosen as potential starting points for BFS; $TH$: threshold for hub nodes' degree; $c_{max}$: the max number of node in an island; $l_{islands}$: list of islands' nodes and hubs info; $P2$: parallel factor
         \STATE {\bfseries Outputs:} None (modify $l_{islands}$ in place)
         \STATE $v_{global} = \{\}$
            \WHILE{$|task| > 0$}
                \FOR[\textcolor{purple}{ \scriptsize \# distributed across $P2$ engines}]{$p=0$ {\bfseries to}  $P2$ {\bfseries in parallel} }
                    \STATE Pop $\{hub_o$,$a_o\}$ from $task$ \{\textcolor{purple}{\scriptsize\# Pop the hub and one of its neighbors}\}
                    \STATE $v_{local}$=$\{ a_o \}$, $h_{local}$=$\{hub_o\}$, $query$=$0$, and $count$=$1$, if $a_o$ is not a hub
                    \WHILE[\textcolor{purple}{\scriptsize  \# if there exist unexplored nodes}]{$query$ $\neq$ $count$} 
                        \STATE $node_o$ = $v_{local}[query]$
                        \FOR{$n$ $\in$ $node_o.neighbors$}
                            \IF[\textcolor{purple}{\scriptsize \# hub node or not}]{ $n.degree < TH$}
                                \IF[\textcolor{purple}{\scriptsize\# $n$ locally explored by engine p}]{$n \in v_{local}$ }
                                    \STATE  Skip neighbor $n$
                                \ELSIF[\textcolor{purple}{\scriptsize\# not explored by other engines}]{$n \notin v_{global}$}
                                    \STATE $count$ $+=$ $1$
                                    \STATE Append $n$ to $v_{local}$ and $v_{global}$\\
                                    \textcolor{purple}{\scriptsize \hspace{0.1em}\# if exceeding the max number of nodes in an island} 
                                    \STATE {\bfseries if} $|v_{local}| > c_{max}$ break {\bfseries while(line 8)}
                                \ELSE[\textcolor{purple}{\scriptsize\# already explored by other engines}]
                                    \STATE remove $v_{local}$ from $v_{global}$
                                    \STATE break {\bfseries while(line 8)} 
                                \ENDIF
                            \ELSE[\textcolor{purple}{\scriptsize\# else it's a hub node}]
                                \STATE Add $n$ to $h_{local}$
                            \ENDIF
                        \ENDFOR
                        \STATE $query$ $+=$ $1$
                    \ENDWHILE
                    \STATE Append  ($v_{local}$, $h_{local}$) to $l_{islands}$ 
                \ENDFOR
            \ENDWHILE
        \end{algorithmic}
        \end{algorithm}
    \end{minipage}
\end{figure}

\begin{figure}[t] 
\centering
\includegraphics[width=3.3in]{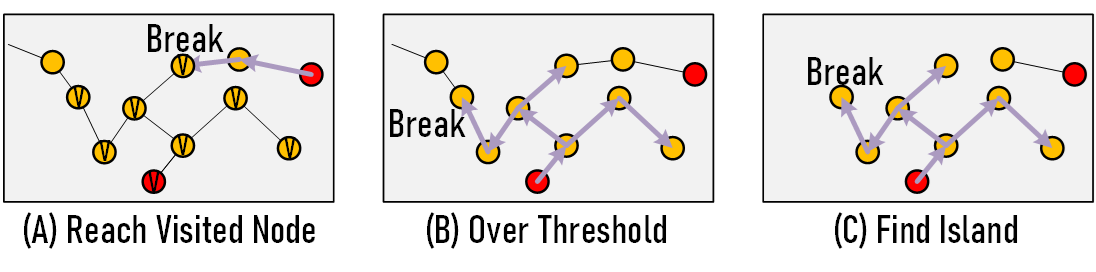}
\vspace{-0.15truein}
\caption{Task break conditions of TP-BFS engine.}
\vspace{-0.13truein}
\label{break_condition}
\end{figure}

To accelerate BFS and reduce off-chip accesses to adjacency data, redundant search must be avoided. To achieve this, the Island Locator keeps a record of the IDs of nodes visited by all TP-BFS engines during a certain round in \textcolor{black}{a global visit list, $v_{global}$.} When a BFS engine reaches a node labeled as visited in the global, but not in the local visited list, it knows that this region has been \textcolor{black}{searched previously by other engines.} The BFS engine then drops this task and waits for the next \textcolor{black}{(line 20 of Algorithm \ref{al:4}} and Figure~\ref{break_condition}(A)). 

As shown in the Algorithm \textcolor{black}{\ref{al:4} (line 14)}, the TP-BFS engine checks that a node is not on the global visit list before adding it to the local visited list and incrementing the visited node counter.
If the counter is over the threshold (expected maximum island size), the engine drops the task and waits for a new one (\textcolor{black}{(line 17)} and  Figure~\ref{break_condition} (B)). 

When all neighbors of the initial node have been scanned, the engine checks whether the query pointer value equals the counter value. This indicates that all nodes have been searched, that TP-BFS is done without reaching the island size threshold, and that an island is found. If this happens, the engine sends the connection information of this island to the Island Consumer and requests a new task from the Task Generator (see Figure~\ref{break_condition} (C)). Otherwise, it accesses the adjacency list of the node pointed to by the query pointer and explores all its neighbors. This process continues until one of the three task-break conditions is triggered.

\subsubsection{Architecture}

\begin{figure}[t] 
\centering
\includegraphics[width=3.4in]{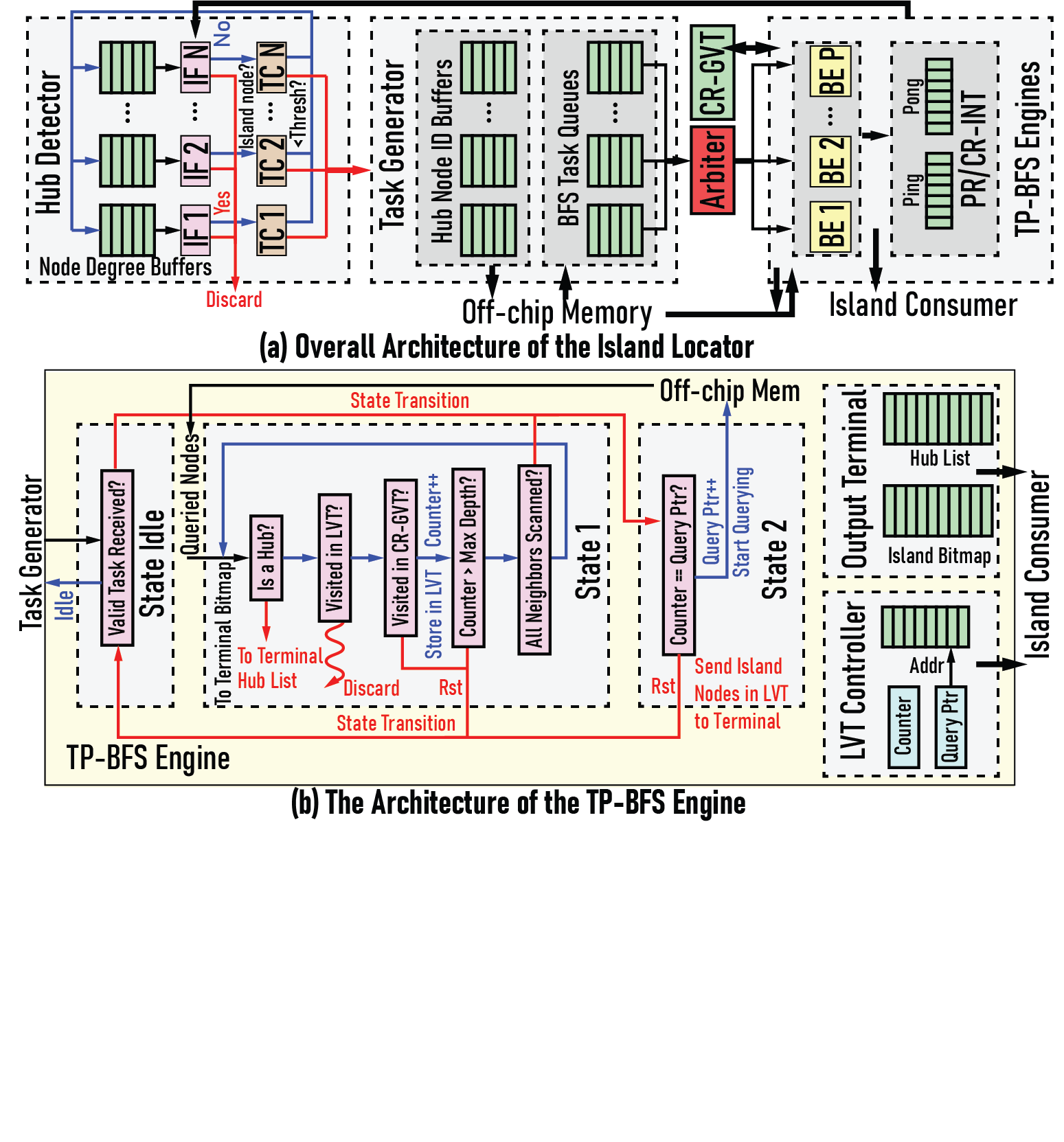}
\vspace{-1.2truein}
\caption{Simplified architecture of Island Locator.}
\label{fig:II_locator}
\vspace{-0.15truein}
\end{figure}

Figure~\ref{fig:II_locator} shows the architecture support of Island Locator. The blue part in Algorithm~\ref{al:1} is realized as the Hub Detector. The degree information of the nodes are distributively stored in the Node Degree Buffers which are implemented with loop-back FIFOs. The number of FIFOs determines the parallelism of the hub detection (P1 in Algorithm~\ref{al:1}). The Island Node Filters (IFs) check whether the nodes popped out of the FIFOs are among the island nodes in the previous round \textcolor{black}{(line 6 in Algorithm~\ref{al:2})} by checking Island Node Table (PR-INT). If yes, these nodes are discarded; if no, the comparators check whether the nodes are hubs \textcolor{black}{(line 8 in Algorithm~\ref{al:2})}. The nodes recognized as hubs are sent to a hub buffer which is also implemented with multi-bank FIFOs; while the remaining nodes are sent back to the Node Degree Buffers for the next round detection. 

The TP-BFS Task Generator realizes \textcolor{black}{Algorithm~\ref{al:3}}. Once a hub node is popped out of the hub buffer, Task Generator will start to access its adjacency list from global memory. The node IDs of the neighbors in the list received by Task Generator and the ID of their hub form task tuples which are cached in TP-BFS Task Queues. These tasks are further assigned to idle TP-BFS engines.

The simplified architecture of TP-BFS is shown in Figure~\ref{fig:II_locator}(b). We implement the orange block of Algorithm~\ref{al:1} as a three-stage Finite State Machine (FSM). At Stage 0, TP-BFS is ideal and sends requests of new tasks to Task Generator. When the new task arrives, TP-BFS engine moves to Stage 2 from Stage 0 with Query Pointer pointing to row 0 of the Local Visited Table (LVT) and the Island Node Counter as 1. In general, at Stage 2, the engine first checks whether the query pointer value and the Island Node Counter are the same. If yes, the engine finds an island, forwards the data stored at the output terminal to the Island Consumer, and records island nodes in CR-INT. Otherwise, the engine accesses the adjacency list of the node pointed by the Query Pointer from global memory, adds one to the Query Pointer, and then moves to Stage 1. At Stage 1, the TP-BFS engine records the newly-arrived adjacency information in the island bitmap buffer at output terminal and tries to find unvisited nodes by scanning the nodes included in this information each per cycle. If the node being scanned is not a hub and also not visited, the Island Node Counter increases by one. If the counter value overpasses the maximum number of nodes for an island, the engine is reset to Stage 0. Otherwise, the engine keeps scanning the list and at the end moves to state 2. 



\subsection{Island Consumer}

Island Consumer follows Island Locator and conducts combination and aggregation of islands and hubs. Before introducing the algorithm and architecture adopted in Island Locator, we first discuss where the redundant calculation of aggregation comes from. 

As islands contain nodes with strong internal connections, it is highly likely that multiple nodes have more than one shared neighbors in which case the aggregation result of these shared neighbors can be reused multiple times with one-time calculation. Figure~\ref{motivation_redundant} uses the graph structure and adjacency matrix of a typical island as a motivational example. As mentioned in Section 3.1, the adjacency matrix of an island includes all connections between island nodes and the connections with hubs. During island processing, all these connections are evaluated. The example graph has seven island nodes from `a' to `g' and one hub `H'.

\textcolor{black}{As shown in Figure~\ref{motivation_redundant}(A1), nodes d,e,f, and g are the shared neighbors of nodes b and c; as an undirected graph, nodes b and c are also the shared neighbors of nodes d, e, f, and g. For the aggregation of nodes b and c, the feature vectors of nodes d, e, f, and g need to be accumulated twice; to process nodes d, e, g, and f, the feature vectors of nodes b and c need to be accumulated four times. If the feature vector length is L, these accumulations take $16 \times L$ operations. However, if we can pre-calculate the accumulation results and reuse them, then only $10 \times L$ operations are needed. Figure~\ref{motivation_redundant}(A2) shows how the pre-calculated results are reused during aggregation. We add two virtual nodes whose feature vectors are the accumulation results {d,e,f,g} and {c,b} in the graph. We connect them to the real nodes according to the accumulation requirements. During the aggregation phase, the pre-aggregated feature vectors are forwarded directly to the target nodes, so that the accumulation from shared neighbors is only done once. Note that in the original graph without island-based locality enhancement, these nodes are highly scattered, as is the processing of their aggregation operations. It is therefore prohibitive to find and prune these repeated operations from shared neighbors at runtime. Through I-GCN, the nodes with strong interconnects are largely clustered, which makes redundancy removal feasible.}

\subsubsection{Algorithm}

We first introduce the calculation methodology of the Island Consumer. Once the information of an island is forwarded to the Island Consumer, it assigns that information to a PE which is waiting for new calculation tasks. The information includes island node IDs, hub node ID, the local adjacency bitmap, the round IDs, and etc. The PE performs first combination and then aggregation reusing the same MAC units.

The PE starts the combination of all hub and island nodes by first accessing their input feature vectors from global memory and then conducting PULL-based combination. Different from conventional combination, the Island Consumer conducts pre-aggregation at the completion of the combination of every k node. Specifically, after the combination results of k nodes have been calculated, we sum them up and use the results in aggregation and so skip the redundant operations. k can be customized.

\begin{figure}[!t] 
\centering
\includegraphics[width=3.3in]{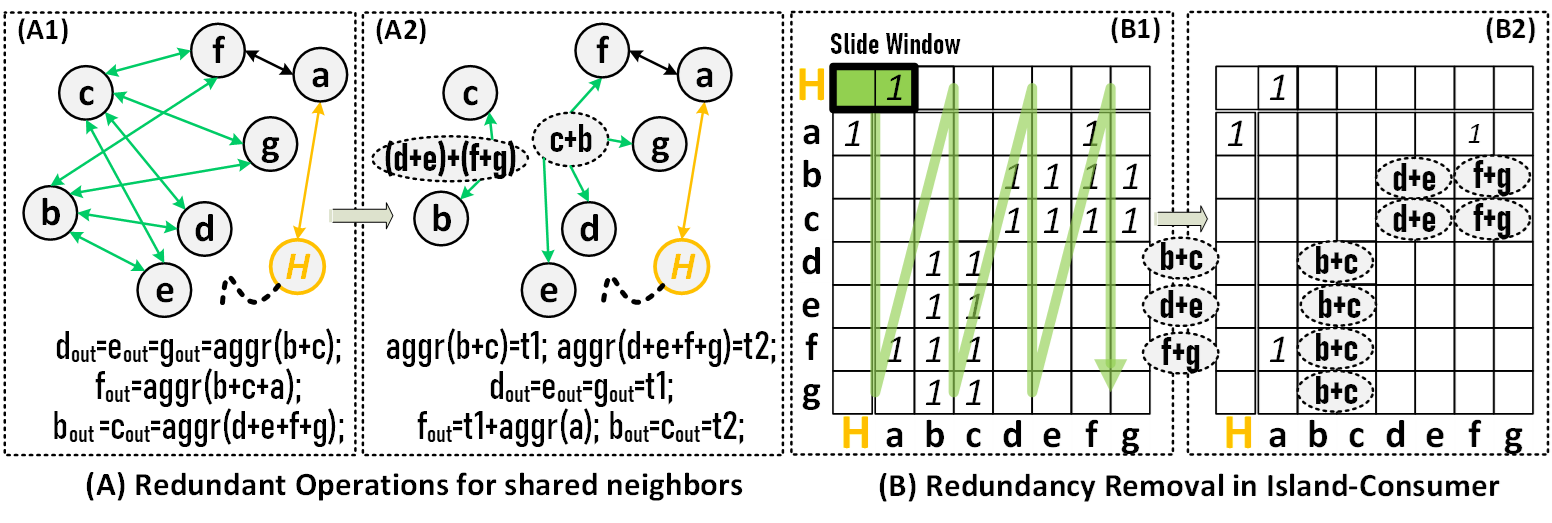}
\vspace{-0.10truein}
\caption{\textcolor{black}{Redundancy removal of a typical island.}}
\vspace{-0.20truein}
\label{motivation_redundant}
\end{figure}

When the combination results of all nodes in an island and their pre-aggregated results are ready, the PE starts aggregation by scanning the local adjacency bitmap. Figure~\ref{motivation_redundant}(B) shows how the Island Consumer uses these results in aggregation with redundancy removal. The scan starts from the top-left corner of the bitmap and slides towards the bottom-right following the green trace. The size of the scan window is $1 \times k$ where k is the number of nodes whose feature vectors are pre-aggregated during the combination phase. If the number of non-zeros covered by the sliding window is less than $k/2$, it is more efficient to accumulate the feature vectors of nodes that are connected (column id of non-zeros under the sliding window) to the node being scanned (row id of the sliding window); otherwise, it is more efficient to subtract the feature vectors of nodes that are not connected (column id of zeros under the sliding window) from the pre-aggregation results. The Island Consumer can automatically pick the one that demands the fewest operations.

Figure~\ref{motivation_redundant}(B) gives an example. For clarity, k is set to 2. The scan starts once the combination of all nodes and all pre-aggregations of adjacent k nodes finish. For each scan, if both bits are non-zeros, i.e. the nodes under-scan are the common neighbors of node-H and node-a, instead of redoing the accumulation, Island Consumer will directly use the pre-aggregation result, saving one vector addition operation. After the first two columns are scanned, Island Consumer proceeds to columns b and c with the combination and pre-aggregation results of node-b and node-c. After the entire bitmap is scanned, both GraphCONV's aggregation and combination of the island have been completed. 

\subsubsection{Architecture}

\begin{figure}[t!] 
\centering
\includegraphics[width=3.4in]{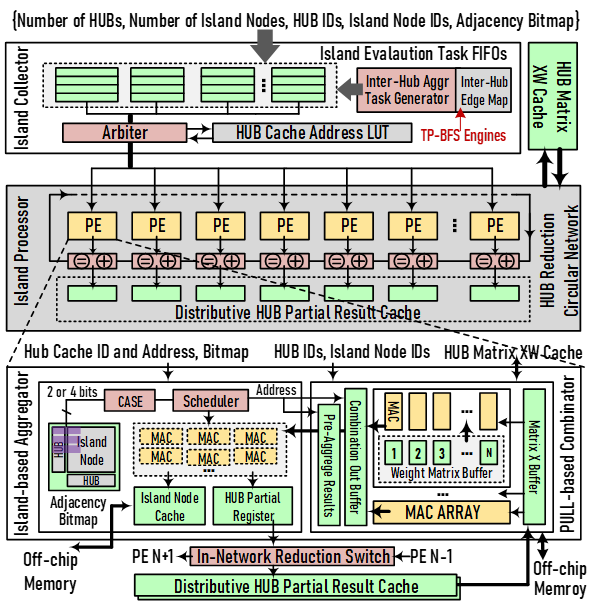}
\vspace{-0.2truein}
\caption{Simplified architecture of Island Consumer.}
\vspace{-0.18truein}
\label{arch_consumer}
\end{figure}

The architecture of Island Consumer is illustrated in Figure~\ref{arch_consumer}. The island-related information sent from Island Locator is received by Island Collector and then stored in the distributed memory of the Island Evaluation Task if the incoming island is not redundant. The major information of each island task includes: numbers of hubs and island nodes, hub node IDs, island node IDs, adjacency bitmap, and the round ID. The arbiters in island collector prefetch evaluation tasks every clock cycle and forward them to the idle PEs. Once a PE receives an evaluation task, it performs the aggregation and combination. At the end of evaluation of each island, the PE produces complete output features of all island nodes and partial results of the output features of hubs. The complete final outputs of island nodes are stored back to the global memory, while the incomplete results of hubs are sent to the corresponding bank of a HUB Partial Result Cache through the ring-based reduction network to update the corresponding partial sums calculated previously. To obtain complete aggregation results of hub nodes, it is necessary to evaluate not only the islands but also the inter-hub connections, which are not included in islands. To perform the aggregation of inter-hub connections, Island Collector maintains an inter-hub edge map based on the information provided by TP-BFS engines during island locating, generates inter-hub aggregation tasks based on the edge map following Push-Outer-Product execution order, and inserts the tasks into the evaluation task queues. More details about the creation of inter-hub edge map and the generation of inter-hub tasks are omitted due to the space limit. Once all islands and inter-hub tasks are evaluated, the complete hubs' results are obtained.

\begin{figure*}[t]
\centering
\begin{minipage}{.61\textwidth}
\noindent
  \includegraphics[width=1.02\linewidth]{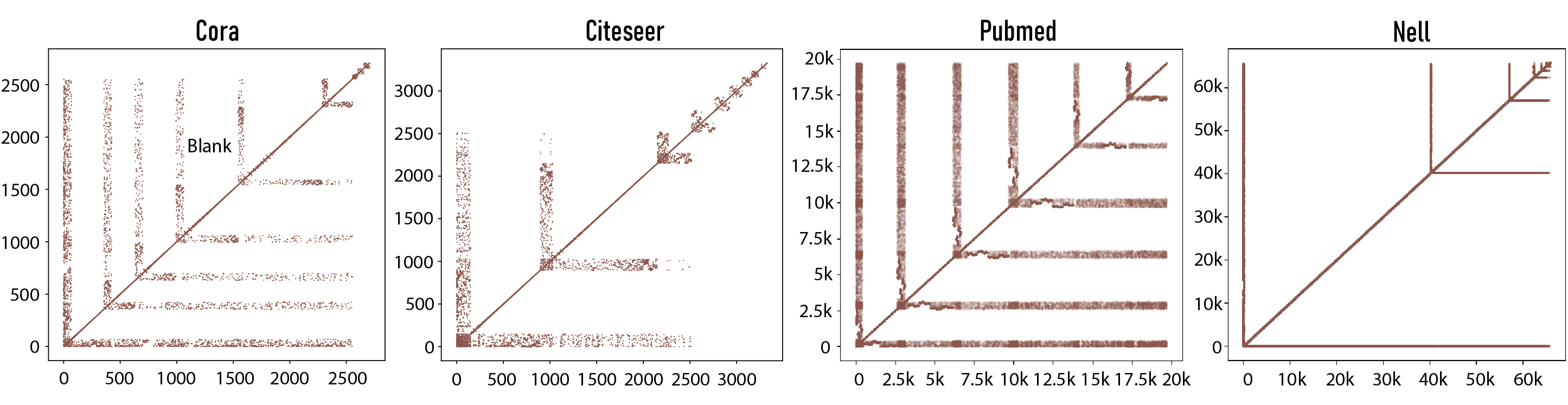}
  \vspace{-0.21truein}
  \captionof{figure}{Islandization effect on Cora, Citeseer, PubMed, and NELL. \textcolor{black}{Space between L-shapes is totally blank.}}
  \label{fig:islandization}
\end{minipage}%
\hfill
\begin{minipage}{.36\textwidth}
\vspace{0.001truein}
  \includegraphics[width=1.01\linewidth]{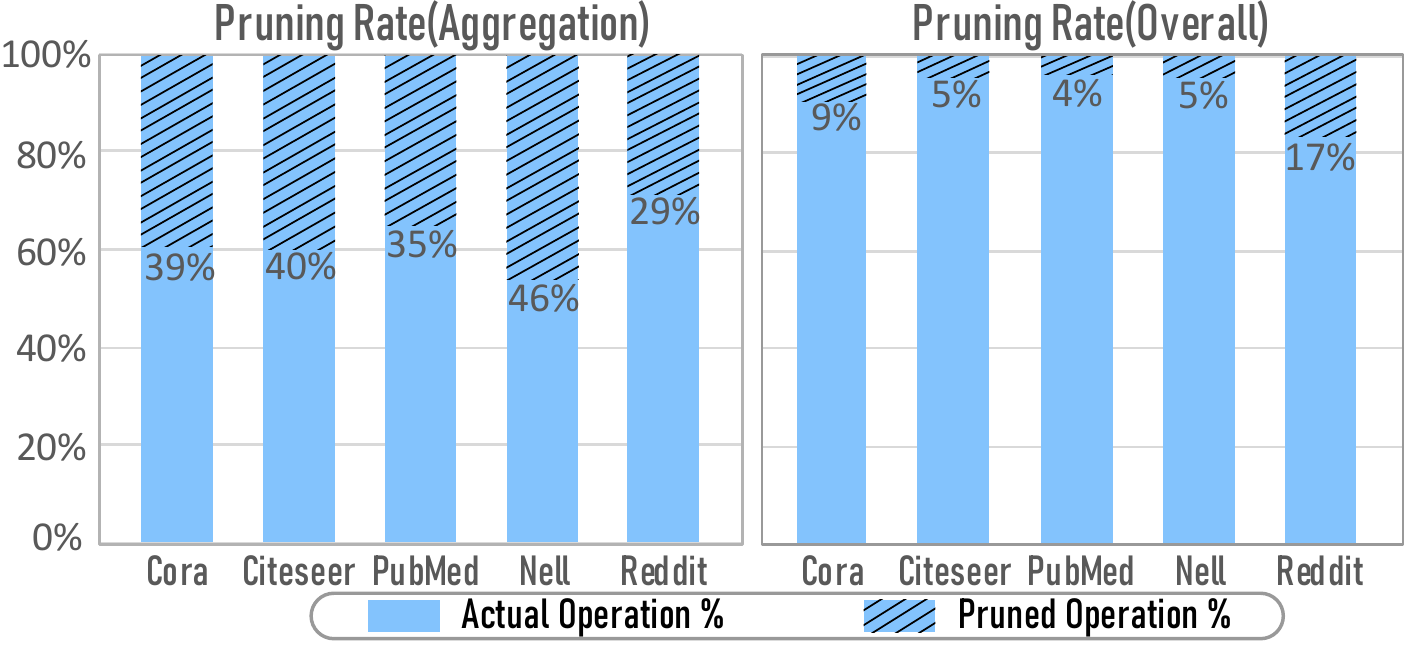}
\vspace{-0.24truein}
  \captionof{figure}{\textcolor{black}{Pruning rates with redundancy removal.}}
  \label{fig:pruning_rate}
\end{minipage}
\vspace{-0.10truein}
\end{figure*}

As shown in Figure~\ref{arch_consumer}, all PEs are connected in a ring network. The partial results of hub nodes are distributively stored in the multi-bank HUB Partial Result Cache (DHUB-PRC). Each bank of DHUB-PRC is attached to one PE. Note that at the first appearance of each hub, the Island Collector will map it to an unused row in a certain bank. The bank ID and row address are attached to the hub before its island evaluation task is assigned to a PE. This bank ID and row address will be fixed and reused when this hub appears again in the future tasks. At the end of each island or inter-hub task processing, the PE will check the hubs' bank IDs. If the bank ID matches its PE ID, the partial results will be used to update the partial sum stored locally, otherwise it will be forwarded to the corresponding banks through the ring network. 

To accelerate the partial sum update of hubs and reduce communication, the ring network is equipped with the support of in-network reduction. Particularly, the switch at each entry of the ring network will check the hub IDs of the partial results sent from local PE and from its left neighbor. If both are valid and the same, they are reduced at the entry and the new result will be sent to the right neighbor at the next cycle. This in-network reduction design is widely used in smart-NIC architectures, so we do not elaborate it due to page limit.

The simplified architecture of each PE is illustrated at the bottom of Figure~\ref{arch_consumer}. The right part is the module for PULL-based combination. As mentioned in Section 3.3.1, this module accesses feature data of island nodes based on their IDs from off-chip. These data are accessed only once as all connections of these island nodes will be processed during the execution of this island task. Weight matrix is distributively stored on-chip in Weight Matrix Buffers of PEs if possible. During calculation, non-zeros of the same node will be sequentially broadcast to the same row of MAC units and are multiplied with the corresponding row of weight matrix. The partial combination results of the same row are locally accumulated. Once a certain node is completely calculated, the row of MAC units starts to work on the next node if applicable. If the node is a hub, the resulting combined features will be stored in HUB Matrix XW Cache for the reuse in future island and inter-hub tasks. Once the combination and pre-aggregation of all nodes are done, the right module will be deactivated and the left one will be activated to start aggregation. The left module is designed based on the calculation method introduced in Section 3.3.1. The purple blocks in the figure are the scan windows. To avoid pipeline bubbles in the aggregation module, the scans with all zeros need to be skipped. To achieve this, we scan multiple rows in parallel at each cycle and only forward the scans with non-zeros to FSMs, i.e. CASE and Scheduler modules in the figure, which access the required pre-aggregation and/or combination output results from the combination module and assign the aggregation task to idle MAC units.

\section{Evaluation}

\subsection{Experiment Setup}
We evaluate I-GCN's latency, energy efficiency, off-chip data movement, and hardware resource utilization with a Stratix 10 SX FPGA. We evaluate I-GCN on three different models -- GCN, GraphSage, and GIN -- using five datasets commonly used in GCN acceleration research \cite{geng2019uwb,yan2020hygcn,liang2020engn}. These include Cora (CR), Citeseer (CS), Pubmed (PM), Nell (NE), and Reddit (RD). With respect to network structures, existing systems use various configurations: GCNs and GraphSage have two layers, while GIN has three layers. For GCNs and GraphSage, EnGN and AWB-GCN use the configurations reported in the original algorithm papers \cite{kipf2016semi}, while HyGCN uses its own configuration of 128 hidden channels for all datasets. 
Here we label GCN and GraphSage with original configurations, ``GCN-algo'' and ``GS-algo'' and label the ones used in HyGCN, ``GCN-Hy'' and ``GS-Hy''. For GIN, HyGCN is the only work that uses it in evaluation. We evaluate all these models and compare with the corresponding existing work.

To better demonstrate the efficiency of the proposed on-the-fly algorithm-architecture co-design for islandization, we compare I-GCN with 6 existing lightweight graph reordering algorithms (Section 4.5). These algorithms are realized on an Intel Xeon Gold 6226R CPU with 64 threads. For cross-platform comparison, we compare I-GCN with prior art GCN accelerators (HyGCN \cite{yan2020hygcn}, AWB-GCN \cite{geng2019uwb}), prior art SpMM accelerator (SIGMA \cite{qin2020sigma}), NVIDIA V100 GPU, RTX8000 GPU, Intel E5-2683-V3 CPU, and Intel E5-2680-V3 CPU. The CPU and GPU results are based on PyTorch Geometric (PyG) \cite{fey2019fast} and Deep Graph Library (DGL) \cite{wang2019dgl}.


\begin{figure}[t!] 
\centering
\includegraphics[width=3.3in]{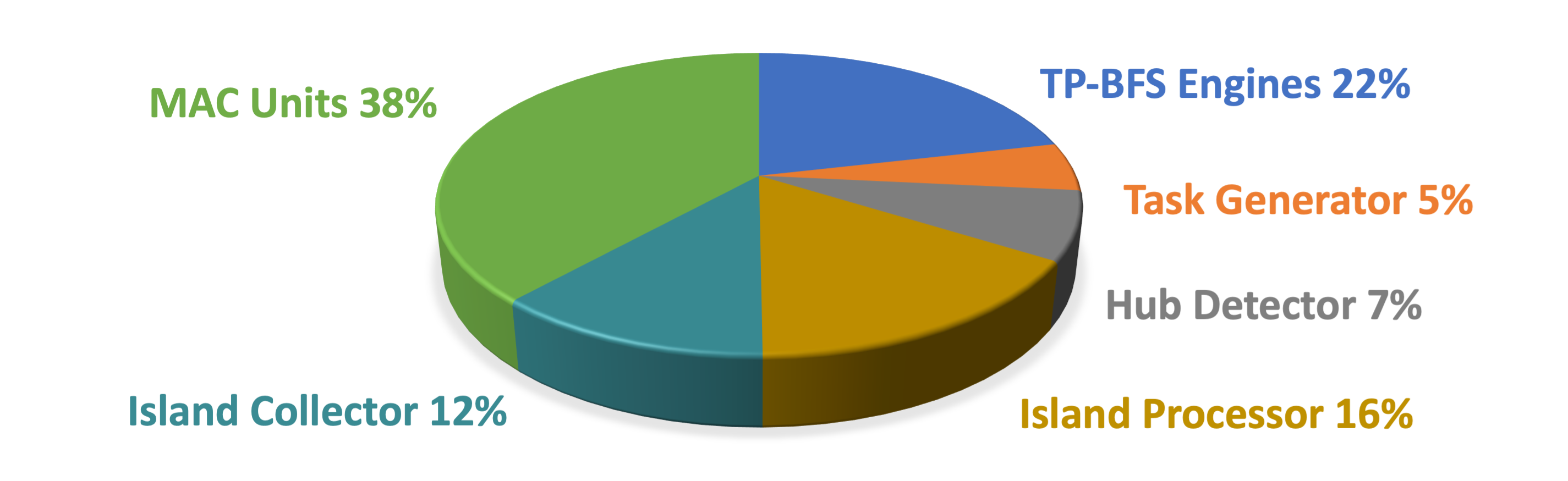}
\vspace{-0.15truein}
\caption{Hardware consumption breakdown of I-GCN.}
\vspace{-0.10truein}
\label{hardware_consumption}
\end{figure}

\subsection{Islandization Effect}

We first evaluate the efficiency of the islandization algorithm. Figure~\ref{fig:islandization} shows the effects of the proposed islandization method on the adjacency matrices of four real-world graphs with various statistics including size, sparsity and non-zero distribution. Among them, NELL is believed to be the most difficult to process due to its extremely high sparsity and imbalanced distribution\cite{geng2019uwb}. As shown in Figure~\ref{fig:islandization}, for all these four datasets, our islandization method is able to optimally cluster all non-zeros to the anti-diagonals and L-shaped clusters within several rounds. The islandization effect on NELL looks even more significant than the other datasets, since its adjacency matrix is the sparsest and its graph has more significant component structures.

\subsection{Island-based Redundancy Removal}

Here we evaluate the effects of shared-neighbor-aware redundancy removal in the Island Consumer. Figure~\ref{fig:pruning_rate} shows the operation pruning rates during aggregation phase. The Island-based Aggregator is able to skip, on average, \textcolor{black}{38\%} of aggregation operations. These operations are all redundant and are for the aggregation among shared neighbors. The removal of these operations is lossless. Note that in combination-first calculation, aggregation phase takes \textcolor{black}{23\%} operations on average. Therefore, \textcolor{black}{9\%} of operations of the entire processing can be eliminated without accuracy loss. Meanwhile, the theoretical latency is lowered by \textcolor{black}{9\%}.

\subsection{Hardware Consumption}

Figure~\ref{hardware_consumption} illustrates the hardware resource usage breakdown of an I-GCN with 4K MAC units and 64 TP-BFS Engines. In order to show comparable breakdowns with ASIC implementation, we normalize the usage of LUTs and Flip-Flops to the number of Adaptive Logic Modules (ALMs) which is the basic component in Intel FPGAs. The Island Locator accounts for 34\% of the entire accelerator. The Island Consumer accounts for the remaining 66\%. Note that in practical FPGA implementations, the MAC units are normally instantiated with DSP slices. Here we normalize the usage of DSP slides to ALMs for a clearer area breakdown.

\begin{figure}[!t] 
\centering
\includegraphics[width=3.4in]{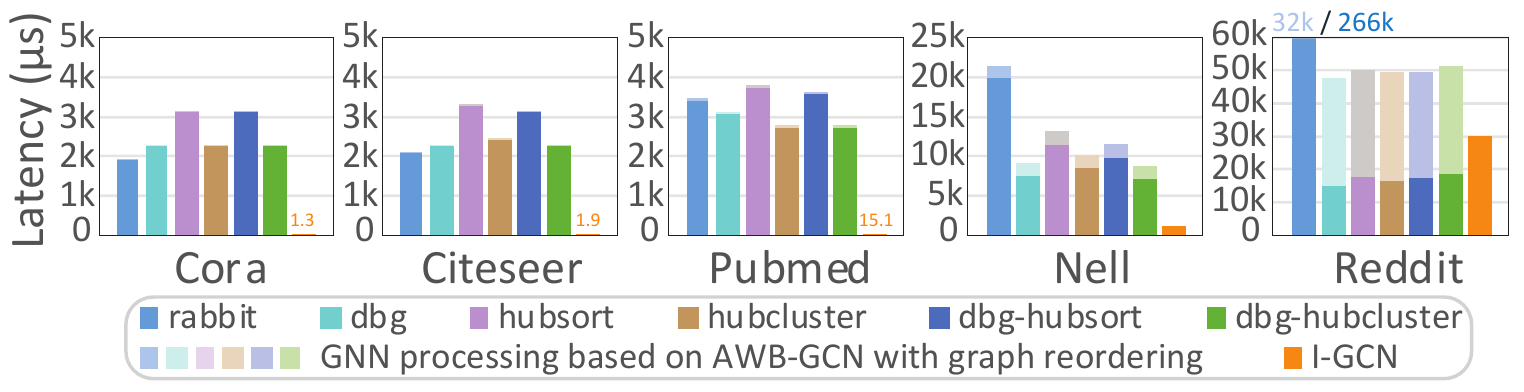}
\vspace{-0.25truein}
\caption{\textcolor{black}{Latency of I-GCN vs AWB-GCN + light weight reordering algorithms.}}
\vspace{-0.15truein}
\label{vslight}
\end{figure}

\begin{table*}[!ht]
\centering
\begin{tabular}{|c|c|c|c|c|c|c|c|c|c|c|c|c|}
\hline
                            & \multicolumn{6}{c|}{\textbf{I-GCN}}                                    & \multicolumn{6}{c|}{\textbf{AWB-GCN}}                                   \\ \hline
                            & \multicolumn{12}{c|}{Device: Intel Stratix 10 SX; Frequency: 330MHz; Num of MACs: 4096}                                                                                    \\ \hline
                            &                           & Cora  & Citeseer & Pubmed & Nell  & Reddit &                           & Cora  & Citeseer & Pubmed & Nell  & Reddit \\ \hline
\textit{\textbf{GCN\_algo}} & \textit{\textbf{Latency}} & 1.3   & 1.9     & 15.1   & 5.9E2 & \textcolor{black}{3.0E4}  & \textit{\textbf{Latency}} & 2.3   & 4.0      & 30     & 1.6E3 & 3.2E4  \\ \hline
\textit{\textbf{}}          & \textit{\textbf{EE}}  & 7.1E6 & 3.7E6    & 5.3E5  & 1.3E4 & 3.5E2  & \textit{\textbf{EE}}  & 3.1E6 & 1.9E6    & 2.5E5  & 4.1E3 & 2.1E2  \\ \hline
\textit{\textbf{GCN\_Hy}}   & \textit{\textbf{Latency}} & 8.2   & 12.9     & 1.1E2  & 1.2E3 & \textcolor{black}{4.6E4}  & \textit{\textbf{Latency}} & 17    & 29       & 2.3E2  & 3.3E3 & 5.0E4  \\ \hline
                            & \textit{\textbf{EE}}  & 9.6E5 & 6.0E5    & 8.1E4  & 7.5E3 & 2.2E2  & \textit{\textbf{EE}}  & 4.4E5 & 2.7E5    & 3.2E4  & 2.3E3 & 1.5E2  \\ \hline
\end{tabular}
\vspace{0.05truein}%
\caption{I-GCN's absolute results of Latency in $\mu s$ and Energy Efficiency (EE) in Graph/kJ.}
\vspace{-0.15truein}%
\label{tb_la}
\end{table*}

\subsection{Comparison with Lightweight Reordering}

\begin{figure}[!t] 
\centering
\includegraphics[width=3.3in]{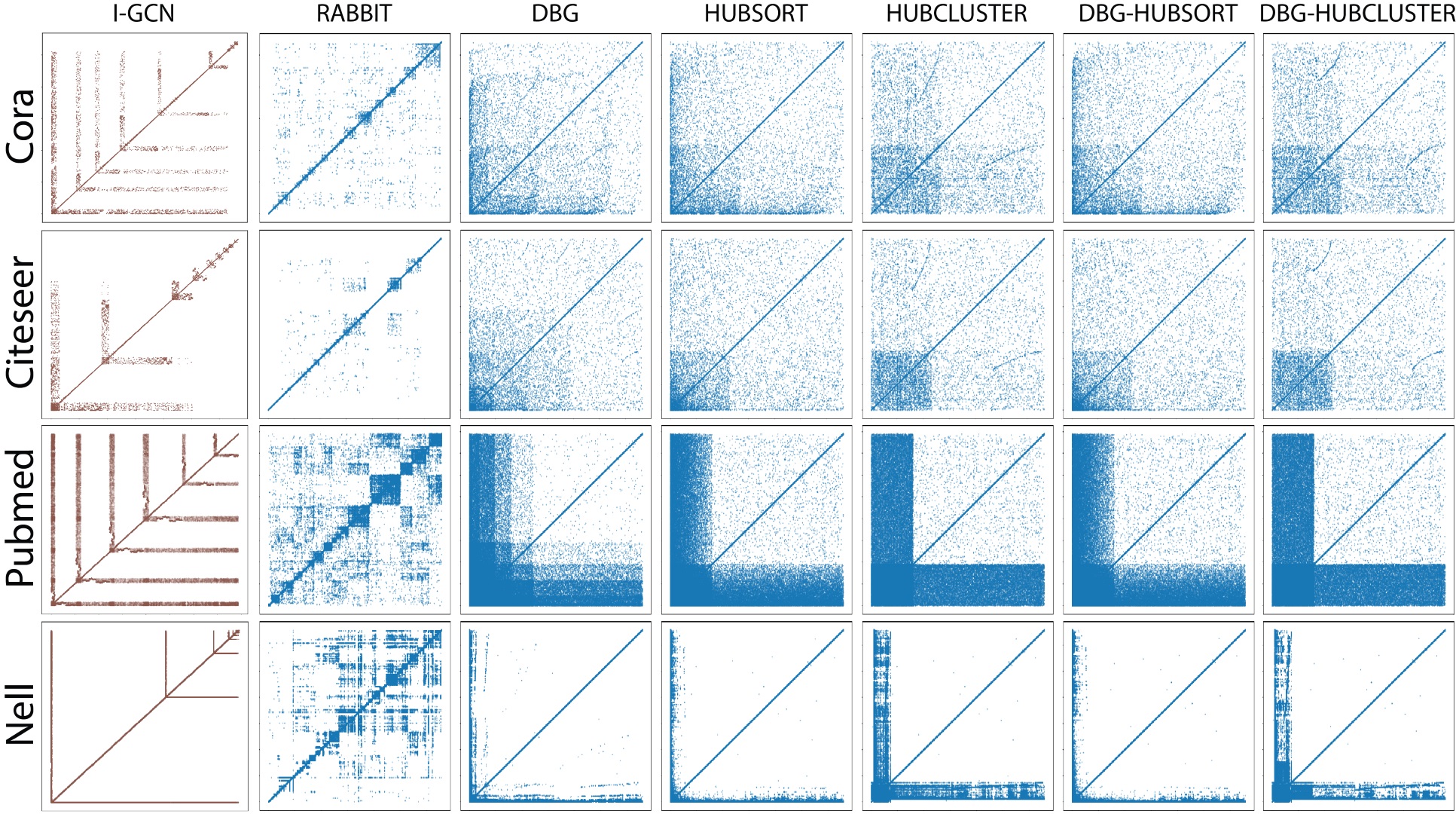}
\vspace{-0.05truein}
\caption{Comparison of non-zero clustering effects}
\vspace{-0.15truein}
\label{cluster_effect}
\end{figure}

To evaluate the benefits of I-GCN over lightweight graph reordering algorithms, we compare I-GCN with 6 baselines. These baselines use 6 traditional lightweight graph reordering algorithms \cite{8573478,7515998,8257937,9041948} for graph preprocessing to enhance locality and then use AWB-GCN \cite{geng2019uwb}, which is a prior art GNN accelerator, to process the reordered graphs. We use these open-source graph ordering codes\cite{9041948} and run them using a high-end Intel Xeon Gold 6226R CPU with 64 threads enabled. There are two findings. First, I-GCN is much faster than the lightweight graph reordering approaches. As shown in Figure~\ref{vslight}, the reordering latency alone can be already higher than the entire I-GCN end-to-end inference latency (more than 100$\times$ for Cora, Citeseer, and Pubmed). And second, I-GCN generates better non-zero clustering. As shown in Figure~\ref{cluster_effect}, I-GCN’s islandization process is able to push all the non-zeros into the L-shaped regions and the anti-diagonal, leaving the remaining area empty. In contrast, the graph reordering methods leave many outlying non-zeros, which introduces significant overhead for their special handling.

\subsection{Cross-platform Comparison}

\subsubsection{Off-chip Data Movement}

\begin{figure*}[!t] 
\centering
\includegraphics[width=7.0in]{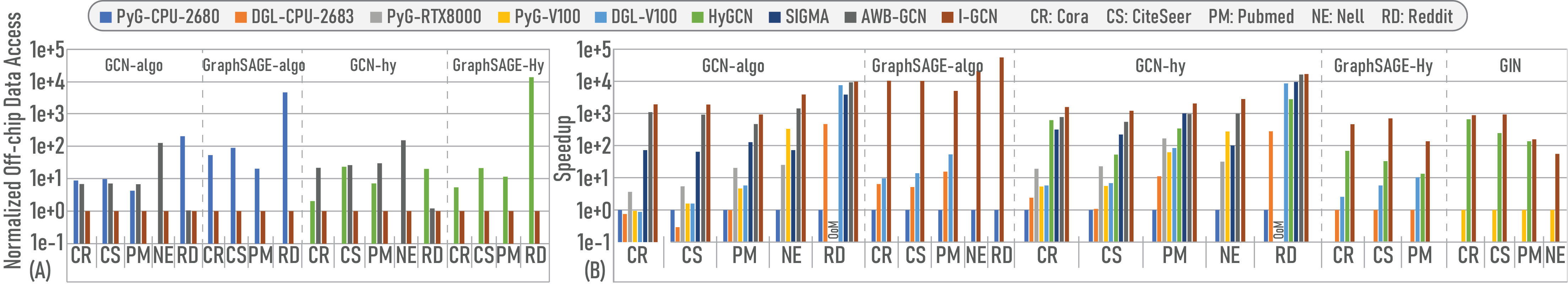}
\vspace{-0.25truein}
\caption{\textcolor{black}{Cross-platform comparison of (A) Normalized off-chip data access and (B) Speedup (end-to-end latency).}}
\vspace{-0.1truein}
\label{longlatency}
\end{figure*}

Figure~\ref{longlatency}(A) compares the normalized numbers of off-chip data accesses of I-GCN with AWB-GCN, HyGCN, and PyG-CPU (Intel Xeon E5-2680-V3) using both GCN-Algo and GCN-Hy. Note that we count the numbers of off-chip accesses assuming that the adjacency matrix and input feature matrix are all stored off-chip at the start of processing. In practice, in the case that the on-chip memory is not fully occupied, these matrices can be partially or even completely stored on-chip to reduce the off-chip bandwidth requirements.

\subsubsection{Latency}

\textcolor{black}{Figure~\ref{longlatency}(B) compares the end-to-end inference latency of I-GCN with SOTA GNN accelerators (AWB-GCN, HyGCN), SOTA SpMM accelerator (SIGMA), PyG-based CPU and GPU, and DGL-based CPU and GPUs. Results show I-GCN provides speedups of $9568\times$ over PyG-based E5-2680-v3 CPU, $1243\times$ over DGL-based E5-2683-v3 CPU, on average $368\times$ over PyG-based GPUs (RTX8000 and V100), $453\times$ over DGL-based V100, $16\times$ over SIGMA, and on average $5.7\times$ over GNN accelerators (HyGCN and AWB-GCN).} Table~\ref{tb_la} lists the absolute results of I-GCN and AWB-GCN. The speedups of I-GCN over AWB-GCN on Reddit are lower than other datasets, as Reddit graph has less significant component structures.  


\textbf{Fairness of evaluation:} HyGCN is an ASIC design which uses 4608 fixed-point MAC units running at 1GHz; AWB-GCN is an FPGA design which uses 4096 floating-point MAC units (running at 330MHz). To provide a fair comparison, the I-GCNs used for evaluation are also equipped with 4096 floating-point MAC units running at 330MHz and use comparable numbers of ALM resources of the same FPGA used by AWB-GCN.

\section{Related Work}

Researchers have designed dedicated hardware architecture to accelerate GCNs \cite{abadal2020computing,chen2020rubik, yan2020hygcn, liang2020engn, geng2019uwb, zeng2020graphact, kiningham2020grip, liang2020deepburning, auten2020hardware,Zhang2021GCOS}.
In \cite{auten2020hardware}, Auten et al., present the first GNN hardware accelerator. By designing four specialized modules -- for graph traversals, dense matrix operations, data scheduling, and graph aggregations, respectively -- the proposed accelerator provides high performance in tackling irregular data movement and intensive computation for GNN inference.
\emph{HyGCN} \cite{yan2020hygcn} is another of the earliest GNN accelerators. With the observation that GCNs are composed of two phases with different computation patterns, HyGCN introduces a hybrid architecture with dedicated modules for aggregation and combination, respectively. 
\emph{AWB-GCN} \cite{geng2019uwb} is another early study of GCN acceleration. It observes that the power-law distribution of the non-zeros in the adjacency matrix results in workload imbalance issues. To solve this problem, the authors propose a workload autotuning technique.
\emph{EnGN} \cite{liang2020engn} uses an unified architecture to accelerate GNNs and adopts a ring-based network to perform aggregation. The results produced by PEs are sent to the network where they are aggregated. Researchers have also designed hardware for training.
\emph{Rubik} \cite{chen2020rubik} proposes an offline graph reordering method to improve data locality.
\emph{GraphACT} \cite{zeng2020graphact} uses heterogeneous platforms with CPUs and FPGAs and uses pre-processing to find and skip redundant operations among two-node shared neighbors. 
\emph{G-CoS}\cite{Zhang2021GCOS} is the first GNN co-search framework for network structure and accelerator architecture. G-CoS can automatically search for the matched GNN structures and accelerators to maximize both task accuracy and acceleration efficiency.

In differentiation from all prior work, the proposed I-GCN is designed to solve the data locality problem fundamentally. As I-GCN reorders graphs using hardware-only solutions, it is compatible with both static and dynamic graphs and both inductive and transductive GNN models. Furthermore, I-GCN finds and skips redundant operations among arbitrary numbers of shared neighbors at runtime.

In graph processing, various reordering algorithms \cite{8573478,7515998,8257937,9041948, chen2021corder,lee2019pre,zhang2017making, wei2016speedup, karantasis2014parallelization} have been proposed for enhancing data locality. The evaluations of six traditional lightweight graph reordering algorithms \cite{8573478,7515998,8257937,9041948} in Section 4.5 demonstrates the high overheads of graph reordering, even for the lightweight ones, which is prohibitive for many real-time GNN inference tasks. There are also many other sophisticated graph reordering algorithms. SlashBurn \cite{lim2014slashburn} is one of them. SlashBurn is able to effectively cluster non-zeros in the adjacent matrices to one of their corners. However, it requires very expensive and complex logic to conduct frequent node degree sorting, graph reconstruction, component size sorting, and node degree updating. Furthermore, SlashBurn is not designed to be parallelized. These make SlashBurn hardware-unfriendly and unsuited for GNN acceleration which normally poses very strict constraints on latency. 


Other architectural and software optimizations have been proposed that improve graph processing efficiency through cache-guided scheduling \cite{8574527,mukkara2017cache}. Among them, HATS \cite{8574527} appears to be the first hardware work that leverages the community structure of graphs without preprocessing. The main goal of HATS, which is tightly integrated with hierarchical cache systems, is to enhance the efficiency of the cache hierarchy for graph processing. In contrast, the islandization in our loosely-coupled I-GCN accelerator aims at clustering non-zeros through community identification for the purpose of extremely fast GNN inference ($\mu$s-level). Also, I-GCN and HATS define components in different ways. HATS detects very coarse-grained and large components by Bounded Depth-First Scheduling (BDFS), while I-GCN locates smaller and more fine-grained components through hub nodes.

Another related topic is hardware acceleration for SpMM \cite{srivastava2020matraptor, srivastava2020tensaurus, hegde2019extensor, pal2018outerspace, qin2020sigma}. Prior art systems include MatRaptor \cite{srivastava2020matraptor}, Extensor \cite{hegde2019extensor}, SIGMA \cite{qin2020sigma}, and Tensaurus \cite{srivastava2020tensaurus}. Although the major kernel of GNN processing is SpMM (see Equation~\ref{eq:gcn_layer}), a high-performance GNN accelerator should be able to fully leverage unique graph features. For example, real-world graphs are extremely sparse, contain communities, and follow the power-law distribution. I-GCN, as a graph-specific architecture, can effectively detect communities from real-world large graphs, and process them more efficiently by avoiding repeated computation for common neighbors. In contrast, SpMM accelerators need to handle all different kinds of sparse matrices. They may behave better for general sparse matrices, but not in those likely to be processed by GCNs.

\section{Conclusion}

This paper proposes a novel hardware accelerator for GCN inference, I-GCN, which significantly improves data locality and reduces unnecessary computation through a new hardware runtime algorithm --- islandization. Islandization finds clusters of nodes with strong internal but weak external connections which yields two major benefits: (1) by processing islands, data can be better reused on-chip which significantly relieves the off-chip bandwidth pressure; (2) there is less redundant computation as aggregation for shared neighbors in an island can be reused. Experimental results show that I-GCN provides speedups over CPUs, GPUs, prior art GCN accelerators of \textcolor{black}{5549$\times$, 403$\times$, and 5.7$\times$}, respectively. 


\begin{acks}
This work was supported by the Compute-Flow-Architecture (CFA) project under PNNL’s Data-Model-Convergence (DMC) LDRD Initiative. The Pacific Northwest National Laboratory is operated by Battelle for the U.S. Department of Energy under Contract DE-AC05-76RL01830.
\end{acks}

\bibliographystyle{ACM-Reference-Format}
\bibliography{refs}

\end{document}